\documentclass[sigconf]{acmart}

\usepackage{subcaption}

\usepackage[colorinlistoftodos,prependcaption,textsize=tiny]{todonotes}
\usepackage{cleveref}

\usepackage{bm}
\usepackage{multirow}

\usepackage{booktabs}
\usepackage{adjustbox}
\usepackage{soul}
\usepackage{amsmath}

\usepackage{microtype}
\usepackage{balance}
\usepackage{xcolor}
\usepackage{tabularx}
\usepackage{colortbl}
\usepackage{paralist}
\usepackage{array}
\usepackage{graphicx}
\usepackage{tcolorbox}
\usepackage{pifont}

\newcolumntype{L}[1]{>{\raggedright\let\newline\\\arraybackslash\hspace{0pt}}m{#1}}
\newcolumntype{C}[1]{>{\centering\let\newline\\\arraybackslash\hspace{0pt}}m{#1}}
\newcolumntype{R}[1]{>{\raggedleft\let\newline\\\arraybackslash\hspace{0pt}}m{#1}}

\usepackage[inline]{enumitem}

\AtBeginDocument{%
  }

\copyrightyear{2025}
\acmYear{2025}
\setcopyright{cc}
\setcctype{by}
\acmConference[SIGIR '25]{Proceedings of the 48th International ACM SIGIR Conference on Research and Development in Information Retrieval}{July 13--18, 2025}{Padua, Italy}
\acmBooktitle{Proceedings of the 48th International ACM SIGIR Conference on Research and Development in Information Retrieval (SIGIR '25), July 13--18, 2025, Padua, Italy}\acmDOI{10.1145/3726302.3730325}
\acmISBN{979-8-4007-1592-1/2025/07}

\begin{document}

\title{A Versatile Dataset of Mouse and Eye Movements on Search Engine Results Pages}

	\author{Kayhan Latifzadeh}
	\orcid{0000-0001-6172-0560}
	\affiliation{%
		\institution{University of Luxembourg}
		\city{Luxembourg}
		\country{Luxembourg}
        }
	\email{kayhan.latifzadeh@uni.lu}

	\author{Jacek Gwizdka}
	\orcid{0000-0003-2273-3996}
	\affiliation{%
		\institution{University of Texas at Austin}
		\city{Austin}
		\country{USA}
        }
	\email{jacekg@utexas.edu}

	\author{Luis A. Leiva}
	\orcid{0000-0002-5011-1847}
	\affiliation{%
		\institution{University of Luxembourg}
		\city{Luxembourg}
		\country{Luxembourg}
        }
	\email{luis.leiva@uni.lu}

\renewcommand{\shortauthors}{Latifzadeh et al.}

\begin{abstract}
We contribute a comprehensive dataset to study user attention and purchasing behavior on Search Engine Result Pages (SERPs).
Previous work has relied on mouse movements as a low-cost large-scale behavioral proxy
but also has relied on self-reported ground-truth labels, collected at post-task,
which can be inaccurate and prone to biases. 
To address this limitation, we use an eye tracker to construct an \emph{objective} ground-truth of \emph{continuous} visual attention.
Our dataset comprises 2,776 transactional queries on Google SERPs, collected from 47 participants, and includes:
(1)~HTML source files, with CSS and images; 
(2)~rendered SERP screenshots; 
(3)~eye movement data; 
(4)~mouse movement data; 
(5)~bounding boxes of direct display and organic advertisements;
and (6)~scripts for further preprocessing the data.
In this paper we provide an overview of the dataset and baseline experiments (classification tasks)
that can inspire researchers about the different possibilities for future work.
\end{abstract}


\keywords{mouse tracking, eye tracking, SERPs, sponsored search}

\maketitle

\section{Introduction}
\label{sec:introduction}

Understanding how users allocate their attention on search engine results pages (SERPs)
can help designers to arrange the elements in more prominent locations 
or to optimize ad placement for better noticeability, for example.
While eye-tracking technologies are commonly used 
in attention studies~\cite{kim2015eye, borys2017eye, ziv2016gaze, sutcliffe2012predicting}, 
mouse movements have been considered a fairly reasonable proxy for user's gaze, 
especially on SERPs~\cite{huang2011no, huang2012user, arapakis2016predicting, leiva2020attentive}.
Unlike eye-tracking, which requires specialized equipment and is typically confined to lab settings, 
mouse movements can be tracked cost-effectively online and at a large scale, 
while users browse within their natural environment~\cite{Leiva13-tweb}.

For a long time, researchers have been interested in decoding mouse movement data 
to infer visual attention on SERPs~\cite{arapakis2016predicting, arapakis2020learning}.
However, previous work~\cite{arapakis2020learning} has relied on post-task questionnaires to obtain ground-truth labels. 
Such self-reporting techniques, although useful, can be biased and inaccurate~\cite{vraga2016beyond, gao2021investigating}.
There is still a lack of studies on user attention on SERPs 
where predictive models of visual attention, based on mouse cursor data, 
are created using \emph{objective} ground-truth labels, such as those derived from eye fixations.
To bridge this gap, we conducted an experiment with 47 participants, 
under controlled lab conditions, using transactional queries on Google Search. 
These type of queries play a crucial role for search engines and online businesses, as they indicate user's purchase intent.
We have collected eye and mouse cursor movements 
together with snapshots of the SERPs in various formats, including source HTML with CSS and images.
Then we have used eye fixations to label areas of interest (AOIs),
with a particular focus on advertisements,
since they represent the most relevant AOIs for the queries studied.

In this paper, we first describe the dataset and all available data types in detail. 
Then, we provide a number of analyses to offer insights into how the data can be used. 
We also present baseline models to demonstrate how to decode user attention via mouse movements.
In summary, we make the following contributions:
\begin{itemize}
    \item A large-scale in-lab dataset comprising mouse and eye movements on SERPs, 
        including self-contained HTML files, screenshots, and scripts for further data preprocessing.
    \item A revisitation of previous research findings about how users interact on SERPs 
        and discussion of new insights derived from this dataset. 
    \item Baseline modeling experiments for assessing users' attention from mouse movements,
        trained on objective labels from eye-tracking fixations.
\end{itemize}

\section{Related work}
\label{sec:relatedwork}

Researchers have shown that mouse movements can reveal a lot of information about the users, 
such as demographics~\cite{leiva2021my}, 
identity~\cite{houssel2024user}, 
satisfaction~\cite{chuklin2016incorporating, liu2015different}, 
preferences~\cite{sadighzadeh2022modeling, schneider2017identifying}, 
experience~\cite{navalpakkam2012mouse, arguello2014predicting}, 
decision making~\cite{katerina2014mouse, bruckner2021choice}, 
next activity~\cite{fu2017your, zhang2024mouse2vec}, 
and visual attention~\cite{arapakis2020learning, leiva2020attentive}. 
In particular, using mouse movements to infer visual attention has shown both potential and challenges,
as discussed next.

Smucker et al.~\cite{smucker2014mouse} investigated the relationship between mouse movements and user attention 
during relevance judgments of web documents. 
For simple tasks, mouse movements did not indicate user attention.
A study by \citet{chuklin2016incorporating} proposed a model 
to jointly capture click behavior, user attention, and satisfaction on SERPs, 
improving predictions as compared to models based on clicks alone. 
\citet{arapakis2016predicting} predicted user engagement 
with the knowledge module of Yahoo SERPs by analyzing mouse movements. 
Their machine learning models outperformed traditional ones.
\citet{liu2017enhancing} found that incorporating mouse movements 
enhanced the accuracy of predicting user clicks and document relevance.
\citet{kirsh2020horizontal} found that horizontal mouse movements, associated with reading direction, 
are related to higher user engagement and interest.
They exploited this observation on a technical-educational website 
to infer user attention during reading tasks~\cite{kirsh2020directions}. 
A recent study by \citet{jaiswal2023predicting} 
used an LSTM model to predict user behavior based on cursor data
and \citet{arapakis2020learning} explored different representations of mouse movements 
to predict user attention to ads on SERPs, although they used self-reported metrics as ground-truth labels.

Previous studies have also investigated the correlation between eye and mouse movements. 
For example, \citet{johnson2012action} showed that mouse movements 
could effectively track visual attention on ads. 
However, their participants were asked to unnaturally move the computer mouse in sync with their eye movements, 
essentially asking them to try to mimic where their eyes were looking on the screen,
similar to \citet{anwyl2022mouseview}.
Another study observed how participants interacted with online news articles, 
using mouse tracking as a proxy for engagement~\cite{arapakis2014understanding}. 
On the other hand, \citet{huang2012user} investigated the relationship between eye and mouse positions during web search, 
focusing on various mouse behaviors such as reading, hesitating, scrolling, and clicking. 
They found that mouse positions do not always match the eye location.
Similar findings were reported by \citet{boi2016reconstructing}.
Finally, \citet{milisavljevic2018eye, milisavljevic2021similarities} 
found that eye-mouse coordination varies with task, scroll speed, and previous eye positions.

Our study builds on this large body of previous work. 
\autoref{table:related-work-datasets} summarizes the most relevant studies, which we briefly describe next. \citet{guo2010towards} collected eye and mouse movement data 
to study the prediction of gaze position from the mouse cursor position on navigational and informational tasks. However, the dataset was not very large and was not publicly shared. 
\citet{huang2012user} also collected both eye and mouse movement data on informational tasks to analyze gaze and cursor alignment. The data has not been released. \citet{mao2014estimating} studied the credibility of user clicks on non-transactional tasks using mouse movement and eye-tracking information. The collected data has not been publically shared. \citet{liu2016predicting} collected both eye and mouse movements to investigate users’ examination behavior.  The dataset is publicly available; it contains screenshots of the SERPs. However no AOI segmentation and no HTML files are provided. \citet{chen2017user} studied user satisfaction prediction using mouse movements. The dataset has never been released. Finally, \citet{arapakis2020learning, leiva2020attentive} provided a publicly available large-scale dataset
of transactional queries but it lacks eye movement data.

In summary, there is no publicly available large-scale datasets of transactional queries that simultaneously provide (1) eye-tracking data, (2) mouse movement trajectories, and (3) SERP HTML source code amenable to programmatic AOI segmentation—particularly at a scale encompassing thousands of log files. To progress beyond the state of the art,
we conducted a large-scale in-lab experiment to collect both mouse and eye movements.
Importantly, we leveraged the Document Object Model (DOM) 
to automatically extract AOIs from thousands of SERPs
and used eye fixations on these AOIs as ground-truth labels 
for inferring visual attention from mouse movements at an high granularity.
To the best of our knowledge, this approach has not been considered and published before. 

\begin{table}[!ht]
    \caption{
        Summary of previous work, sorted by publication year. 
        The `Large' column indicates studies that analyzed more than 1,000 log files.
    }
    \centering
    \small
    \resizebox{1\linewidth}{!}{
    \begin{tabular}{*8c}
        \toprule
        \textbf{Ref.} & \textbf{Transactional} & \textbf{Large} & \textbf{Eye} & \textbf{Mouse} & \textbf{AOIs} & \textbf{HTML} &
        \textbf{Availability} \\
        \midrule
       ~\cite{guo2010towards} & \ding{55} & \ding{55} & \checkmark & \checkmark & \ding{55} & \ding{55} & \ding{55} \\
       ~\cite{huang2012user} & \ding{55} & \checkmark & \checkmark & \checkmark & \ding{55} & \ding{55} & \ding{55}\\
       ~\cite{mao2014estimating} & \checkmark & \ding{55} & \checkmark & \checkmark & \ding{55} & \ding{55} & \ding{55}\\
       ~\cite{liu2016predicting} & \ding{55} & \checkmark & \checkmark & \checkmark & \ding{55} & \ding{55} & \checkmark\\
       ~\cite{chen2017user} & \ding{55} & \checkmark & \ding{55} & \checkmark & \ding{55} & \ding{55} & \ding{55}\\ 
       ~\cite{arapakis2020learning, leiva2020attentive} & \checkmark & \checkmark & \ding{55} & \checkmark & \checkmark & \checkmark  & \checkmark\\
        Ours & \checkmark & \checkmark & \checkmark & \checkmark & \checkmark & \checkmark & \checkmark\\
        \bottomrule
    \end{tabular}
    }
    \label{table:related-work-datasets}
\end{table}

\section{Methodology}
\label{sec:methodology}

\subsection{Participants}
\label{subsec:participants}

\begin{figure*}[!ht]
  \centering
  \includegraphics[width=1\linewidth]{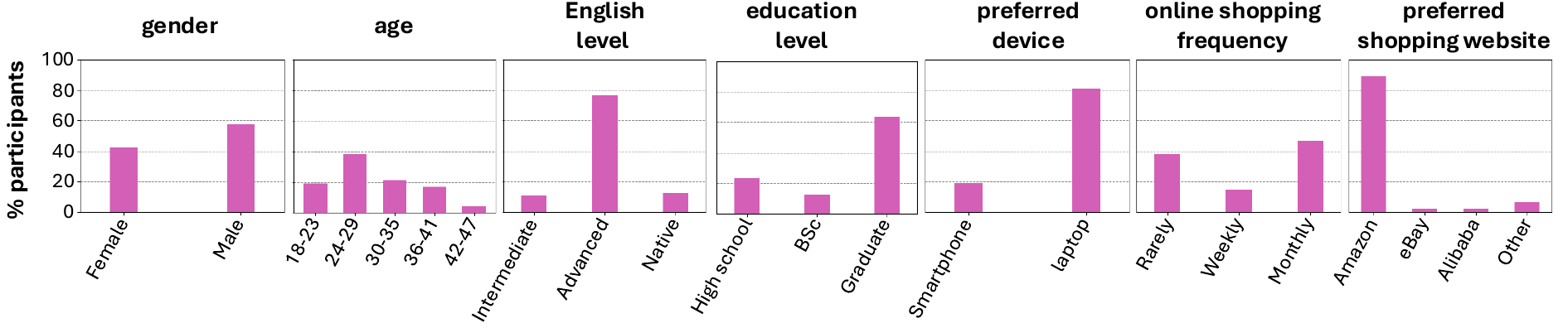}
  \caption{
    Demographics of our user sample ($N=47$ participants).
  }
  \label{fig:demographics-characteristics}
\end{figure*}

Forty-seven participants (27 male, 20 female) were recruited via the University of Luxembourg's mailing lists and flyer advertisements. 
Participants' ages ranged from 19 to 44 years ($M=29.66$, \textit{SD} = 6.46, \textit{Mdn} = 29). 
All participants reported having normal or corrected-to-normal vision and at least intermediate knowledge of English 
(B2 level and above, according to the CEFR\footnote{\url{https://www.cambridgeenglish.org/exams-and-tests/cefr/}}). 
Participants provided written consent and were compensated with 20 EUR. 
The demographic characteristics are shown in \autoref{fig:demographics-characteristics}. 
The Ethics Review Panel (ERP) of the University of Luxembourg reviewed 
and approved the study under ID `ERP 21-055'.

\subsection{Materials}
\label{subsec:materials}

\begin{figure}[ht]
  \centering
  \begin{subfigure}{0.9\linewidth}
    \centering
    \includegraphics[width=\linewidth]{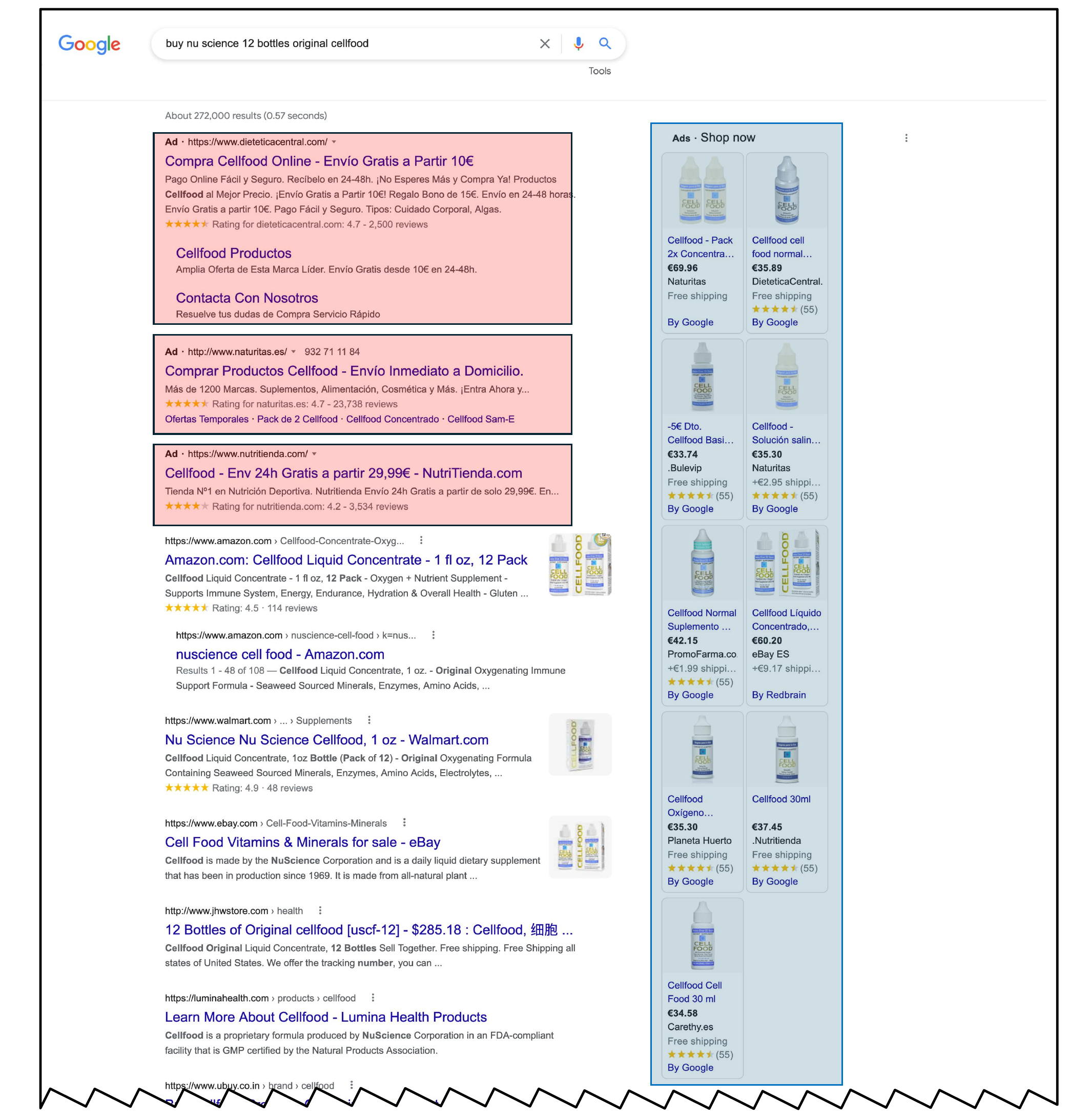}
    \caption{Organic and right-aligned DD ads}
  \end{subfigure}
  \\[1em]
  \begin{subfigure}{0.9\linewidth}
    \centering
    \includegraphics[width=\linewidth]{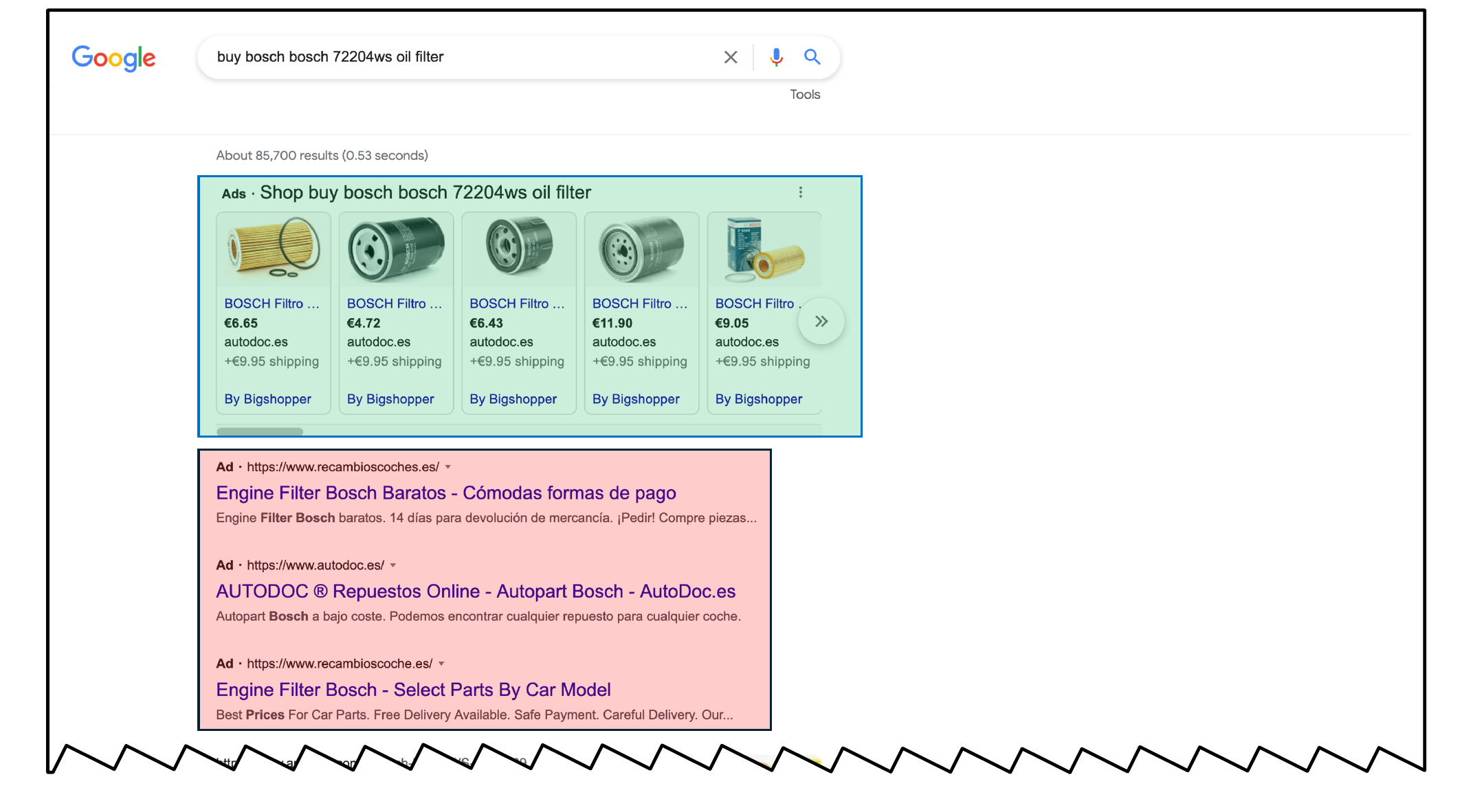}
    \caption{Organic and left-aligned DD ads}
  \end{subfigure}
  \\[1em]
  \begin{subfigure}{0.9\linewidth}
    \centering
    \includegraphics[width=\linewidth]{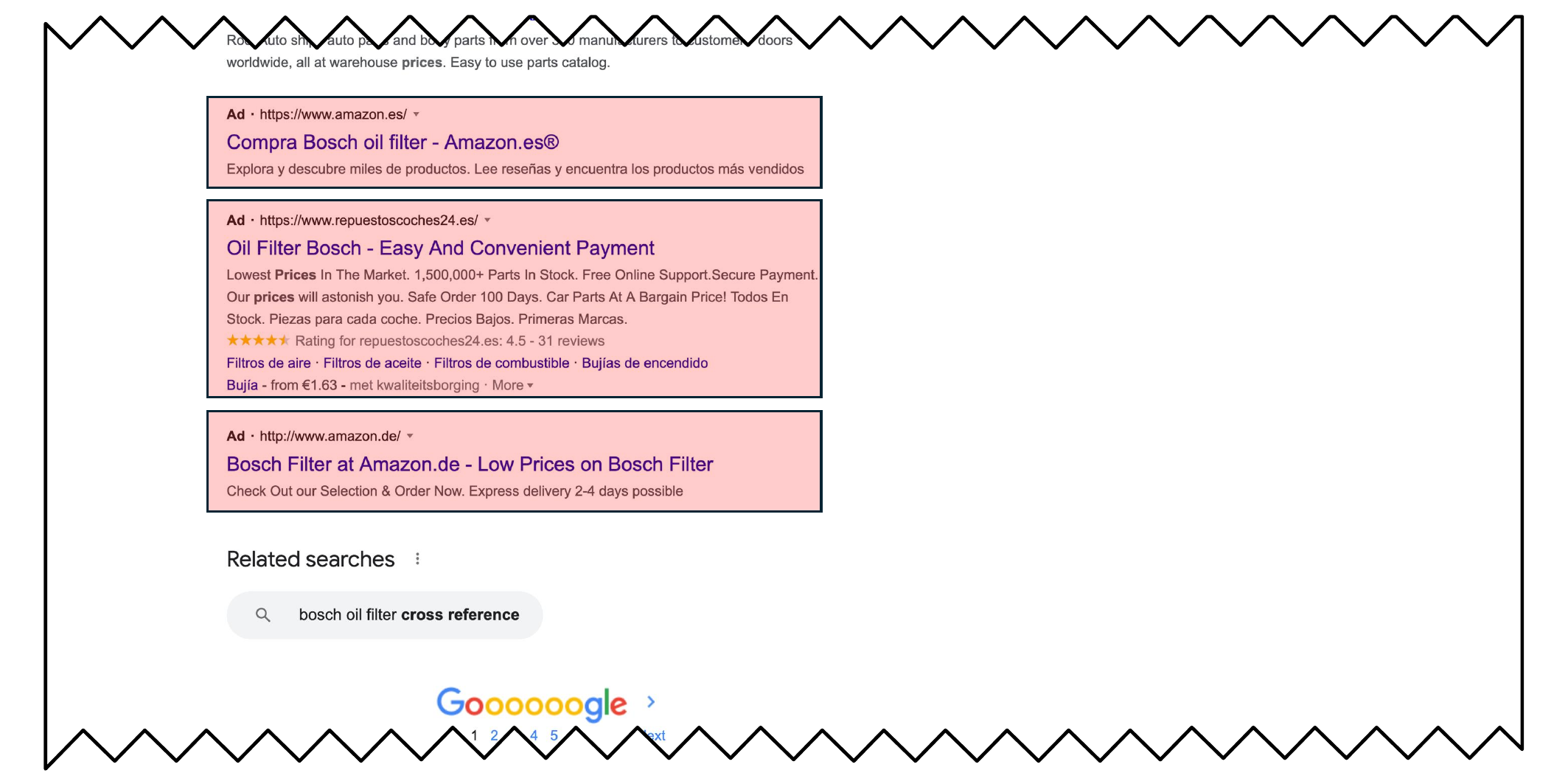}
    \caption{Organic ads}
  \end{subfigure}  
  \caption{Examples of SERPs with ads: organic ads are highlighted in light red, left-aligned DD ads in light green, and right-aligned DD ads in light blue.} 
  \label{fig:serp-ads-example}
\end{figure}

Using the Amazon Product Reviews dataset~\cite{McAuley15_dataset} as our source corpus, 
we generated search queries by prepending ``buy'' to product titles 
and subsequently collected 3,020 unique English-language SERPs. 
Each SERP contained a combination of organic results and direct display (DD) advertisements (\autoref{table:serp-layout-stats}). 
The spatial distribution followed a consistent pattern: 
organic ads appeared either at the top or bottom part of the SERPs, 
while DD ads appeared at the top-left or top-right part.
Representative examples of these SERP layouts are shown in \autoref{fig:serp-ads-example}. 
We randomly distributed the SERPs in 302 batches of 10 SERPs each. 
The initial two batches were designated as familiarization trials, 
allowing participants to learn the experimental protocol.
Each SERP was browsed by one participant only.

\begin{figure}[!ht]
  \centering
  \includegraphics[width=\linewidth]{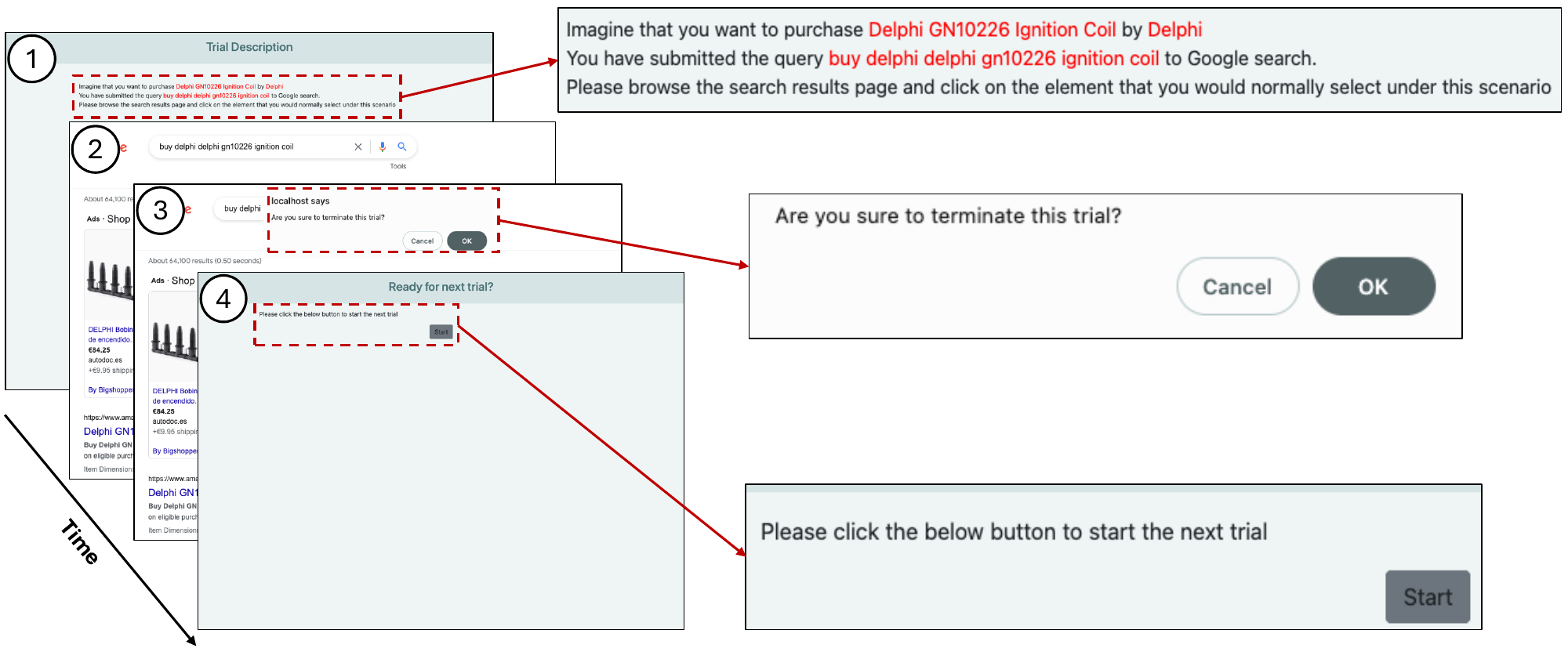}
  \caption{
    Example of an experiment trial. 
    The participant is first informed about the product and its associated query~(1). 
    Then, the participant examines the SERP~(2). 
    If they click on an item, they can decide whether to terminate the current trial or remain~(3). 
    Finally, they can proceed to the next trial whenever they are ready~(4).
  }
  \label{fig:trial-steps}
\end{figure}

\subsection{Apparatus}
\label{subsec:setup}

We developed a web application that prompted participants
to examine the SERPs after a given transactional query, as described in the next section.
We used a 17-inch Dell 1707FP LCD monitor (1280x1024\,px resolution and 60\,Hz refresh rate). 
To minimize on-screen distractions, all SERPs were displayed in full-screen mode.
Eye movements were recorded with a Gazepoint GP3 HD eye tracker, operating at a 150\,Hz sampling rate. 
Mouse movements were recorded with a Dell MS116 mouse 
and the evtrack library.\footnote{\url{https://github.com/luileito/evtrack}}

\subsection{Procedure}
\label{subsec:procedure}

The main experiment consisted of six blocks per participant, each containing 10 trials.
The eye-tracker was re-calibrated before starting each block.
Participants were given at least 1 minute of rest between blocks. 
\autoref{fig:trial-steps} illustrates an example of how each trial is presented to the participants.
At the beginning, participants were provided with a product title and a corresponding query. 
Following a similar setup as the one proposed by~\cite{leiva2020attentive},
participants were instructed to imagine they intended to purchase an item 
after having submitted the given query to Google Search. 
Participants then examined the SERP and had up to 1 minute to explore it 
and make a decision by clicking on the item they would typically choose in that scenario.
They were prompted to confirm their choice. 
Had they not confirmed, they would continue examining for another 1 minute at most.

\section{Dataset description}
\label{sec:dataset-description}

The dataset, which we have named AdSERP, 
is publicly available at Zenodo~\footnote{\url{https://zenodo.org/records/15236546}}.
We also share various preprocessing scripts at GitHub~\footnote{\url{https://github.com/kayhan-latifzadeh/AdSERP}} under an MIT license.

\subsection{HTML files}
\label{sec:html-files}
We provide for each SERP the corresponding HTML source code, self-contained with associated CSS and images;
i.e., everything needed to visualize the SERP is in a single HTML file.
Additionally, we provide one XML log file for each participant's trial,
informing about the corresponding trial and block ID, screen dimensions, window size, and document details.

\subsection{Screenshots}
\label{sec:screenshots}
We also provide rendered full-page screenshots of the corresponding SERPs.  
All screenshots have the same viewport width (1280\,px) but vary in height, depending on the document size. 
The screenshots were captured using Python and the Selenium WebDriver.\footnote{\url{https://www.selenium.dev/}}

\subsection{AOIs}
\label{sec:aois}
We provide a `slot boundaries' file 
containing the bounding boxes of advertisements on the provided screenshots, 
calculated programmatically by traversing the page DOM, 
also using Python and the Selenium WebDriver. 
The ad boundaries are encoded as $(x, y, w, h)$, 
where $x$ and $y$ denote the top-corner coordinates of the ad
(relative to the top-left corner of each screenshot), 
and $w$ and $h$ denote the width and height of the ad, respectively.
Other AOI boundaries can be computed with the scripts we provide to further preprocess the dataset (\autoref{sec:eye-movement}).

\subsection{Eye movements}
\label{sec:eye-movement}

\subsubsection{Pupil Size}
In the format \((t, x, y, p_r, p_l)\),  
where \(t\) represents Unix timestamps in milliseconds,  
\(x\) and \(y\) denote the gaze positions of the eye (relative to the top-left corner of the screenshot) in pixels,  
and \(p_r\) and \(p_l\) indicate the diameter of the eye's pupil in pixels for the right and left eye, respectively.

\subsubsection{Fixations}
In the format \((t, x, y, d)\),  
where \(t\) represents Unix timestamps in milliseconds,  
\(x\) and \(y\) denote fixation positions (relative to the top-left corner of the screenshot) in pixels,  
and \(d\) indicates fixation duration in milliseconds.

\subsection{Mouse movements}
\label{sec:mouse-movement}
Mouse movement data are provided in the format $(t, x, y, e, \text{xpath})$, 
where $t$ represents timestamps, 
$x$ and $y$ denote cursor positions (relative to the top-left corner of the screen), 
$e$ refers to the related mouse event (e.g., scroll, mousemove, or click), 
and $\text{xpath}$ refers to the target element that relates to the event.

\subsection{Preprocessing Scripts and Extra Materials}
\label{sec:eye-movement}
\begin{figure*}[!ht]
  \centering
  \includegraphics[width=\linewidth]{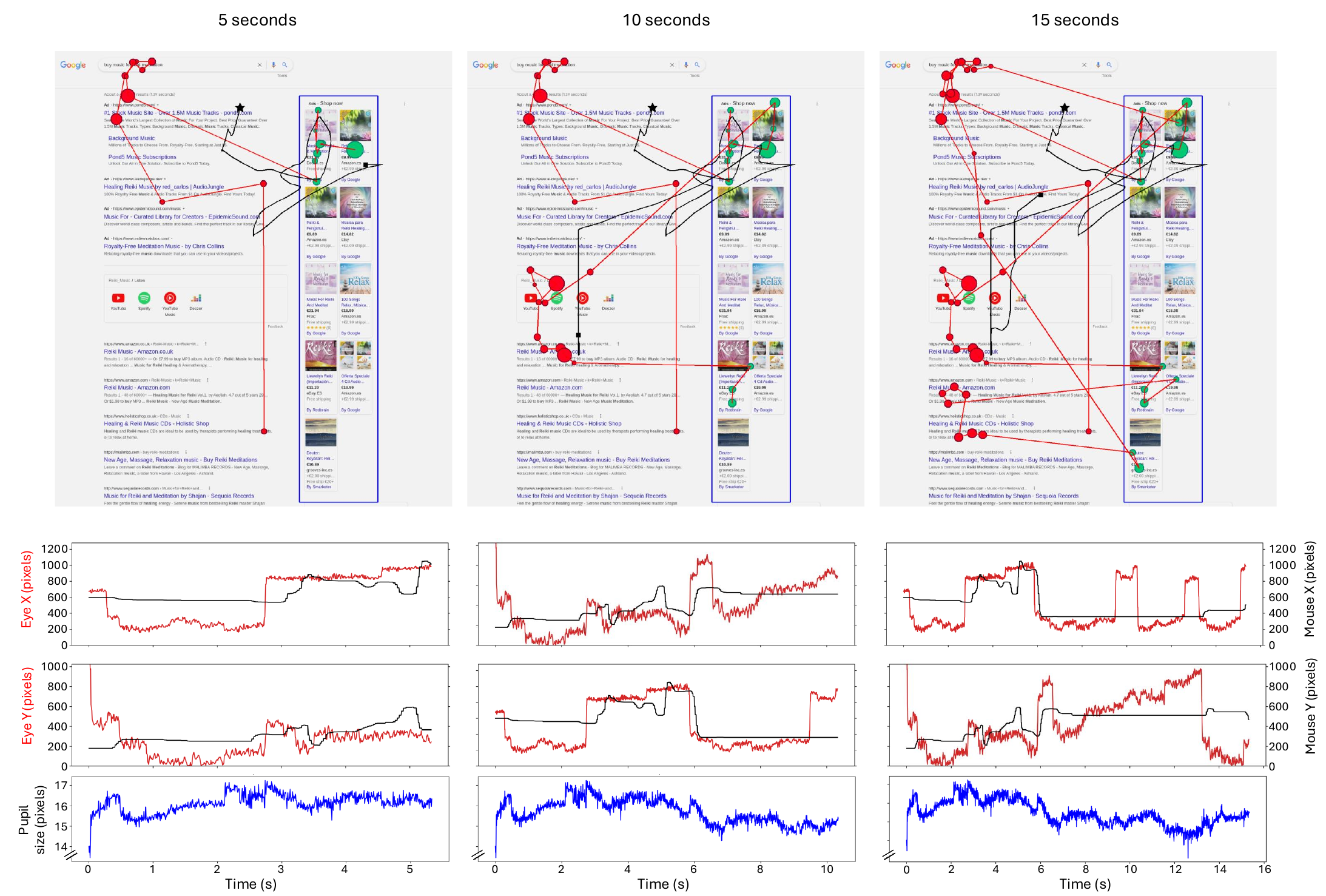}
  \caption{
    Visualization of mouse cursor and eye fixations 
    from a single trial of a participant at three different durations:
    5, 10, and 15 seconds from the start of the trial. 
    The boundary of the right-aligned ad is marked with a blue border. 
    Fixations within the ad boundaries are shown as green circles, 
    while fixations outside these boundaries are represented by red circles. 
    The larger the circle, the longer the duration of the fixation. 
    The mouse trajectory is depicted as a continuous black line, 
    with the start indicated by a star and the end by a square. 
    At the bottom, the mouse and eye screen coordinates, along with pupil size, are plotted for each duration.
  }
  \label{fig:fixation-cursor-examples}
\end{figure*}

We provide Python scripts for loading all the data types in the dataset
(e.g. mouse data, eye data, screenshots, etc.)
\autoref{fig:fixation-cursor-examples} illustrates a visualization of different signals that can be extracted over different durations of a trial.
Additionally, the scripts include examples for calculating different AOIs from the HTML files. 
Moreover, we included a set of visualizations for mouse movements 
proposed by~\cite{arapakis2020learning} in a previous study:

\begin{itemize}
\item \textit{Heatmap}: Mouse cursor influence is calculated using a 2D Gaussian kernel (25 px radius). Overlapping kernels are summed.
 
\item \textit{Trajectories}: Consecutive coordinates are connected by straight lines, with start and end points shown as green and red cursor icons.
 
\item \textit{Colored Trajectories}: Line colors follow a temperature gradient—green for the start and red for the end.

\item \textit{Variable Thickness Trajectories}: Line thickness represents time, with thicker lines at the start and thinner lines at the end.

\item \textit{Colored}, Variable Thickness Trajectories: Combines color gradients and thickness variations.
\end{itemize}

\begin{figure*}[!ht]
  \centering
  \includegraphics[width=\linewidth]{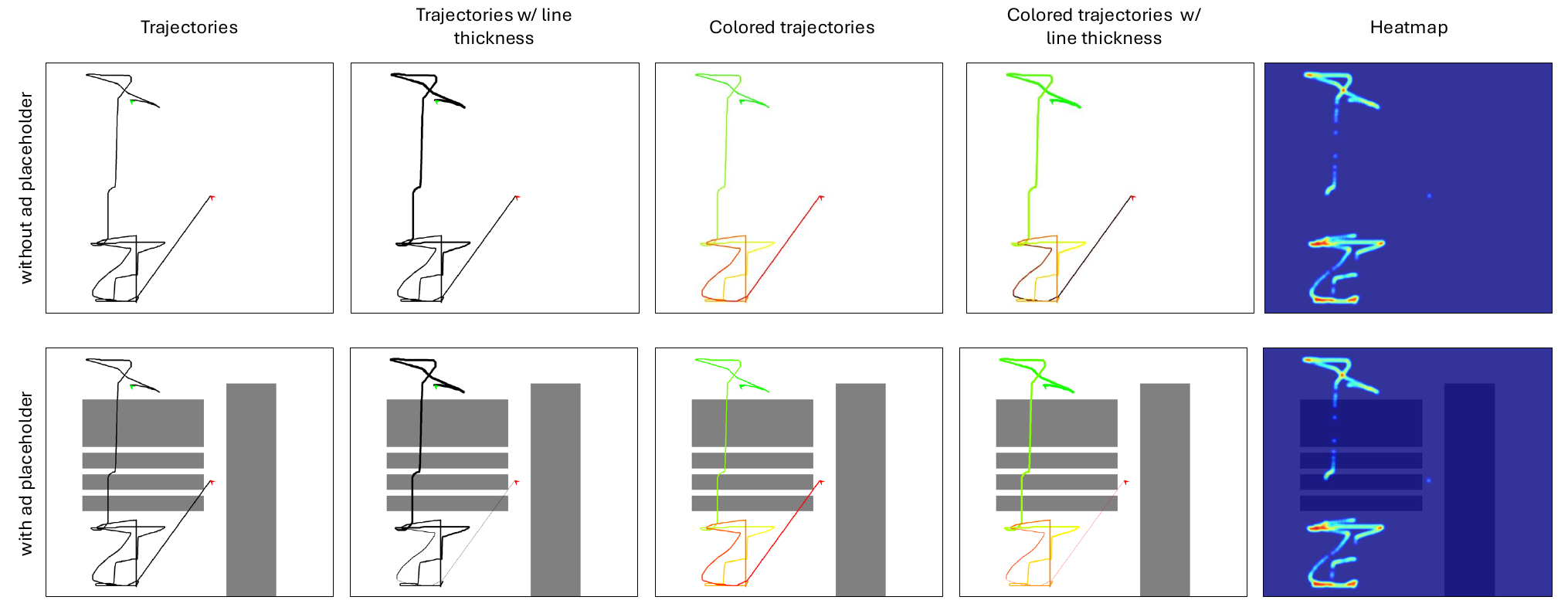}
  \caption{
    Different types of mouse movement visualizations, 
    as proposed by \protect\citet{arapakis2020learning},
    that we also provide.
  }
  \label{fig:representations}
\end{figure*}

We provided these visualizations both with and without advertisement masks (organic and left/right DD ads). 
\autoref{fig:representations} shows an example of these visualizations.
These resources can serve as additional materials for future research experiments. 

\subsection{Screen Recordings}
\label{sec:screen-recordings}
\begin{figure*}[!ht]
  \centering
  \includegraphics[width=\linewidth]{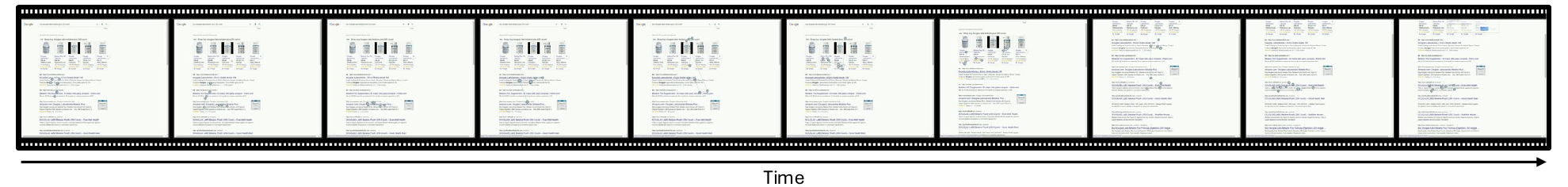}
  \caption{
    Example frames of a screen recording. 
    The gaze locations and the cursor position are overlaid on top of the page.
  }
  \label{fig:screen-recoed-example}
\end{figure*}

We provide screen recordings of each trial. 
These recordings, exported using the Gazepoint Analysis software, 
contain projections of gaze locations on the screen (see \autoref{fig:screen-recoed-example}).

\section{Analysis}
\label{sec:data-analysis}

We analyzed 2,776 SERPs~\footnote{We excluded 44 trials due to malformed log files.} 
featuring various combinations of ad types, 
summarized in \autoref{table:serp-layout-stats}.
The average trial duration was 22.16 seconds (SD = 13.20), with a median of 20 seconds.

In 97.4\% of trials, users completed their task with a single click, as expected, 
while in 2.4\% of the trials a second click was necessary. 
Only 0.2\% of the trials used a third click to finalize their decision.
In 8.61\% of the trials, users finalized their decision by clicking on a left-aligned DD ad, 
3.78\% on a right-aligned DD ad, 
5.19\% on an organic ad.
As observed, the majority of clicks (82.42\%) happened on non-ad elements. 

\begin{figure*}[!ht]
  \centering
  \includegraphics[width=\linewidth]{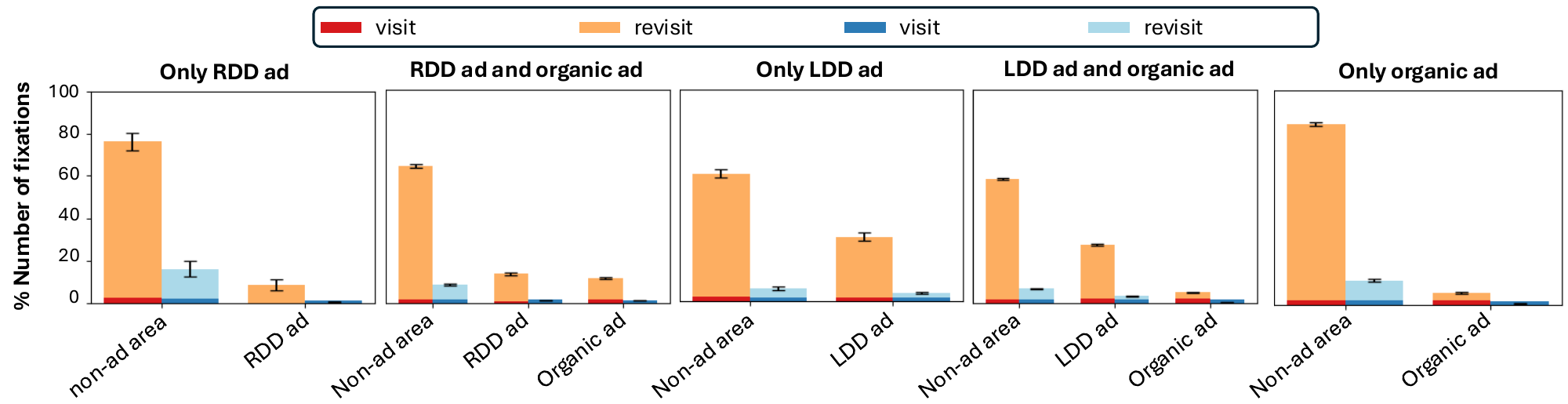}
  \caption{
    Percentage of number of fixations on ad and non-ad areas. 
    Error bars represent standard error of the mean. 
    RDD = right-aligned direct display ad; 
    LDD = left-aligned direct display ad.
    \textit{Visit} refers to the first fixation within an ad 
    while \textit{revisit} refers to fixating back on the same ad after having fixated elsewhere.
  }
  \label{figure:fixation-counts-per-layout}
\end{figure*}

\begin{figure*}[!ht]
  \centering
  \includegraphics[width=0.75\linewidth]{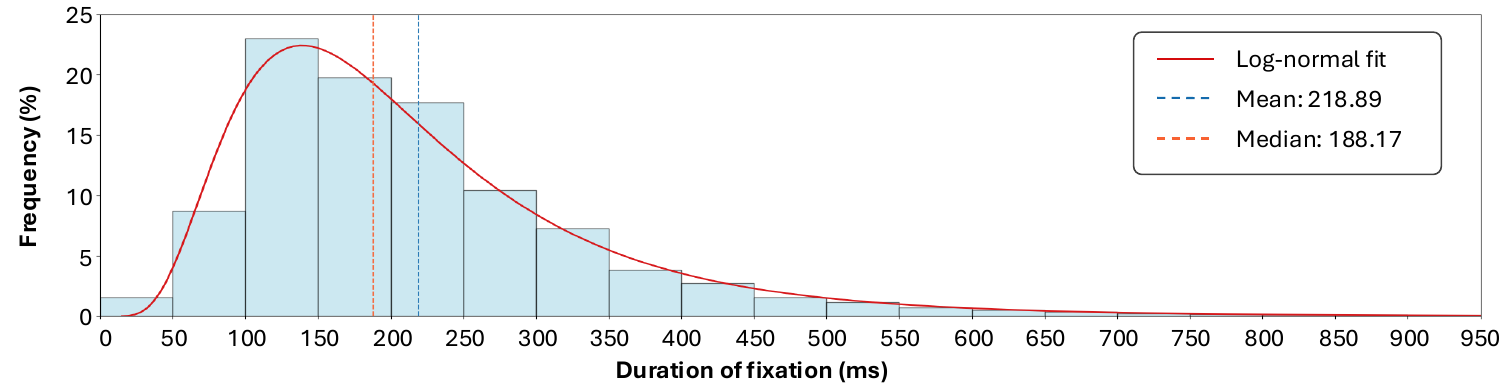}
  \caption{
    Distribution of fixation durations of all participants.
  }
  \label{fig:fixation-duration-histogram}
\end{figure*}

A key measurement provided by the eye tracker is the sequence of fixation points.
We used them to investigate the number of fixations on ad-related and non-ad areas in different SERP layouts 
(see \autoref{figure:fixation-counts-per-layout}). 
We noticed that DD ads tend to capture user attention in much larger proportions than previously known, 
while organic ads receive comparatively less attention.
Another key measurement provided by the eye tracker is fixation \emph{duration}, 
which indicates the level of attention directed at a specific location on the screen. 
Research has shown that longer fixation durations correspond to higher attentional focus~\cite{meghanathan2015fixation}. 
\autoref{fig:fixation-duration-histogram} shows a histogram of fixation durations of all participants.
We can observe a skewed distribution, also reported by \citet{negi2020fixation}.
Similar to previous work~\cite{liu2016predicting, wunderlich2021eye, balatsoukas2012eye}, 
we filtered out fixations with a duration of less than 100\,ms.
The average total number of fixations per trial was 84.42 ($SD=50.61, Mdn=75$).
The average fixation duration was 218.12\,ms ($SD=130.61, Mdn=187$).

\begin{table}[!ht]
    \caption{
        Number of different SERP layout combinations.
    }
    \centering
    \small
    \begin{tabular}{lcc *5c}
        \toprule
        \textbf{SERP layout combination} & \textbf{Number of trials}\\
        \midrule
        Only right-aligned DD ad & 27\\
        Right-aligned DD ad + organic ads & 834\\
        Only left-aligned DD ad & 76\\
        Left-aligned DD ad + organic ads & 1506\\
        Only organic ad  & 307\\
        No ad  & 26\\
        \bottomrule
    \end{tabular}
    \label{table:serp-layout-stats}
\end{table}

\subsection{Mouse and eye coordination}
\label{sec:mouse-eye}

\begin{figure*}[!ht]
  \centering
  \begin{subfigure}{0.32\linewidth}
    \centering
    \includegraphics[width=\linewidth]{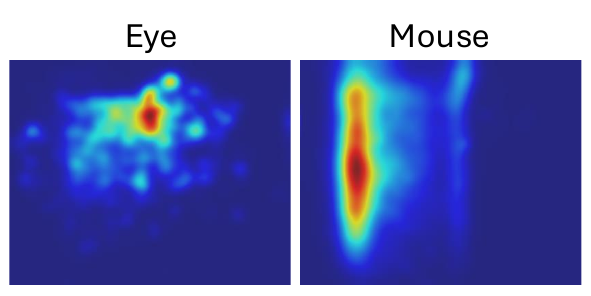}
    \caption{Only organic ads}
    \label{figure:gaze_cursor_heatmap_onlyorganic}
  \end{subfigure}
  \hfill
  \begin{subfigure}{0.32\linewidth}
    \centering
    \includegraphics[width=\linewidth]{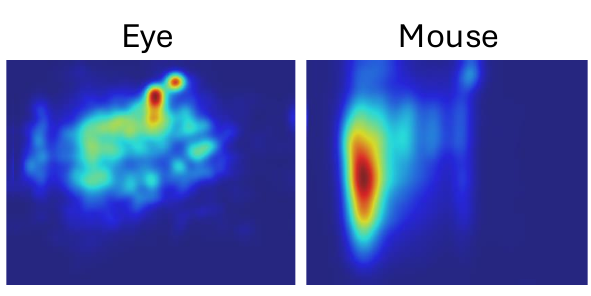}
    \caption{LDD and organic ads}
    \label{figure:gaze_cursor_heatmap_ldd_organic_ads}
  \end{subfigure}
  \hfill
  \begin{subfigure}{0.32\linewidth}
    \centering
    \includegraphics[width=\linewidth]{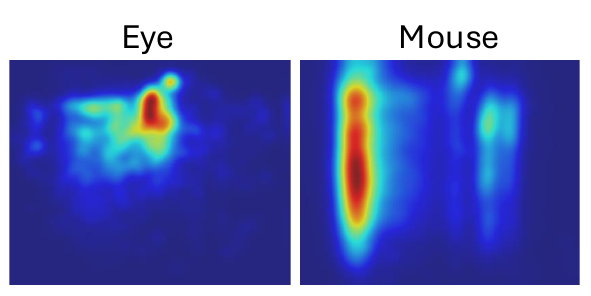}
    \caption{RDD and organic ads}
    \label{figure:gaze_cursor_heatmap_rdd_organic_ads}
  \end{subfigure}
  \caption{
    Heatmaps of eye and mouse locations (1280x1024\,px viewport) for different combinations of ads.
  }
  \label{figure:gaze_cursor_heatmap}
\end{figure*}

\autoref{figure:gaze_cursor_heatmap} provides heatmaps of eye and mouse movements in different SERP layouts.
The mutual information score~\cite{shannon1948mathematical} (higher is better)
is 0.02 for SERPs having right-align DD and organic ads,
0.01 for SERPs having only organic ads,
and 0.06 for SERPs having left-align DD and organic ads.
These differences are not statistically significant,
according to the ANOVA test: $F(2,2644) = 0.81, p = .4451$.
The corresponding Kullback-Leibler divergence~\cite{kullback1951information} (lower is better) 
is 19.89 for SERPs having right-align DD and organic ads,
21.90 for SERPs having only organic ads,
and 17.27 for SERPs having left-align DD and organic ads.
These differences are statistically significant: $F(2,2644) = 11.93, p < .0001$.
Post-hoc pairwise $t$-tests (Bonferroni-Holm corrected) 
revealed that eye and mouse disagree more on SERPs with only organic ads.

\begin{figure*}[!ht]
  \centering
  \includegraphics[width=0.95\linewidth]{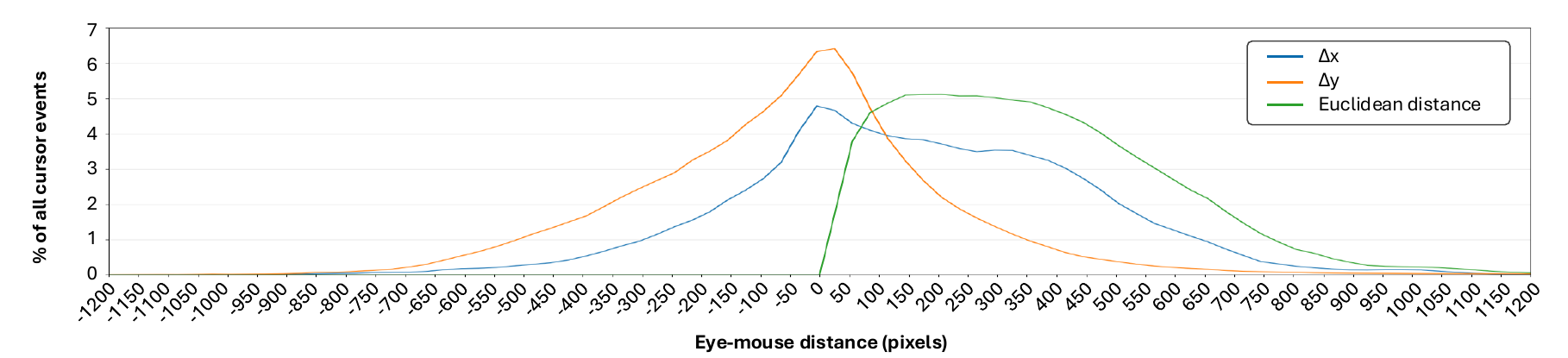}
  \caption{
    Coordination between mouse and eye movements for all SERPs.
  }
  \label{figure:mt_et_distance}
  \end{figure*}

\autoref{figure:mt_et_distance} illustrates the distance between eye and mouse positions.
\(\Delta x\) and \(\Delta y\) are the offsets between eye and the mouse in the X and Y axes, respectively.
In both axes, a peak occurs near 0, where the mouse and eye positions agree the most.
This finding was also observed in an earlier study by \citet{huang2011no}. 
However, in contrast to that study, we found that the distance between eye and mouse is greater along the Y-axis than along the X-axis. 
The mean Euclidean distance is 372.89\,px ($SD=293.78, Mdn=329.83$),
which is almost twice the value reported by \citet{huang2011no} ($M=178, SD=139$).

\subsection{Quantifying attention}
\label{sec:quantifying}

We propose a simple metric to quantify visual attention based on fixations spent on desired AOIs. 
This metric implicitly considered both number of fixations and their duration:
\begin{equation}
\text{Attention}_{\text{trial}} = \frac{\sum \text{Fixation Duration}_{\text{AOI}}}{\sum \text{Fixation Duration}_{\text{total}}}
\label{eq:attention_trial}
\end{equation}
where \( \text{Attention}_{\text{trial}} \) is a number between 0 and 1.

\begin{figure*}[!ht]
  \centering
  \includegraphics[width=0.95\linewidth]{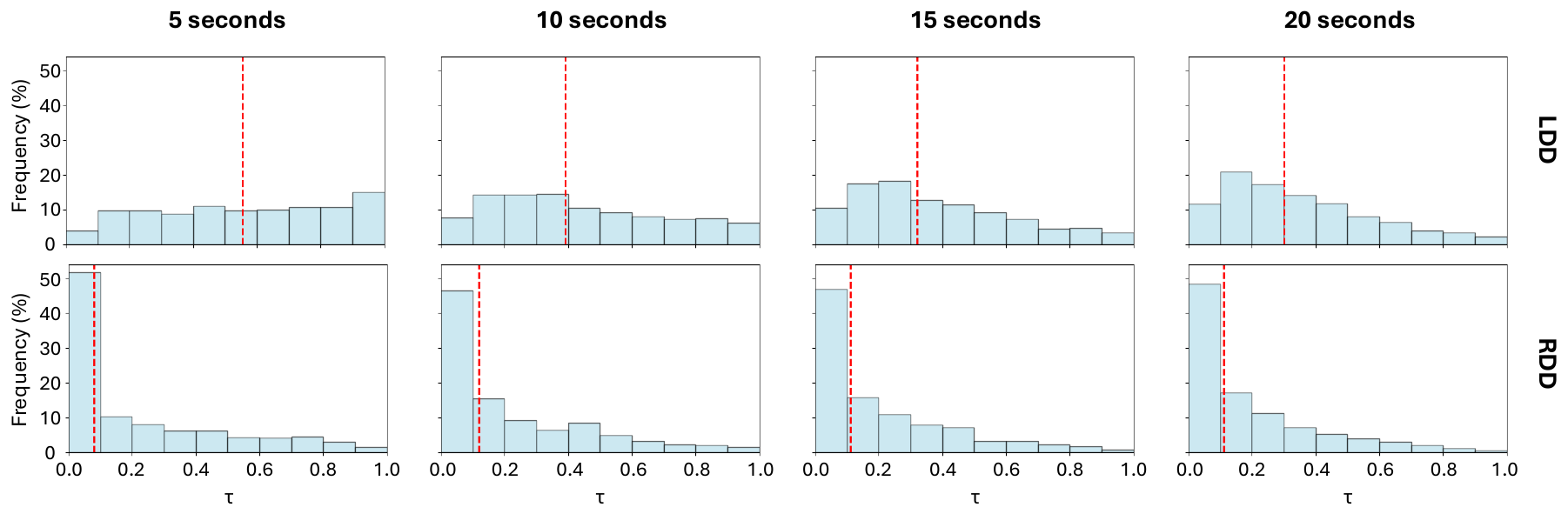}
  \caption{
    Distribution of attention on RDD and LDD ads across all trials. 
    Dashed red lines show the median values.
  }
  \label{fig:attention-ldd-rdd}
\end{figure*}

The amount of attention can also be converted to a binary value 
to determine if an AOI received comparatively sufficient attention or not:
\begin{equation}
\text{label}_{\text{trial}} =
\begin{cases}
1 & \text{if } \text{Attention}_{\text{trial}} > \tau, \\
0 & \text{otherwise}.
\end{cases}
\label{eq:label_assignment}
\end{equation}
where \( \tau \) is a user-defined threshold, typically computed from the collected data, depending on the type of AOI and input duration. 
For example, \autoref{fig:attention-ldd-rdd} illustrates how the attention values are distributed over different durations for DD ads. 
Therefore, the decision on the threshold is a parameter that can be adjusted based on the specific situation or goal of the study.

\section{Data representation and modeling}
\label{sec:representationss-and-models}

We conducted a series of baseline experiments to predict visual attention allocation 
on SERP ads (organic ads and DD ads) based on mouse movements,
inspired by previous work~\cite{arapakis2020learning, leiva2020attentive, leiva2021my, bruckner2021choice}.
Mouse data (multivariate time series) consist of \{$x$, $y$, $t$, $e$\} tuples defined in \autoref{sec:mouse-movement}.
As mentioned in \autoref{sec:data-analysis}, we filtered out fixations with a duration of less than 100\,ms.

We assigned a binary label to each trial, indicating whether the target ad attracted attention, 
as determined by \autoref{eq:attention_trial} and \autoref{eq:label_assignment}. 
For each experiment, we selected the median value of $attention$ for $\tau$.
We also investigated the minimum trajectory duration to predict users' attention,
by considering the first 5, 10, 15, and 20 seconds
(as mentioned in \autoref{sec:data-analysis}, the median trial duration is 20 seconds).
Shorter sequences were zero-padded, while longer sequences were truncated.

\subsection{Input representations}
\label{sec:input-representations}

We consider Mouse2Vec embeddings to train Support Vector Machine (SVM) and $k$-Nearest Neighbors ($k$-NN) classifiers
as well as time series data to train Recurrent Neural Network (RNN) classifiers.
Mouse2Vec~\cite{zhang2024mouse2vec} is a pre-trained model
that receives preprocessed mouse data resampled to 20\,Hz
and returns a 128-dim feature embedding for each 5-second windows of data. 
In addition to these 128 features, we concatenated the bounding box of the desired AOI, 
encoded as $(x,y,w,h)$ tuples denoting the $x,y$ position of the AOI and its size (width and height).
For time series data, at each timestep we consider the mouse position $(x,y)$
and also concatenate the bounding box information.

\subsection{Model architectures}
\label{sec:computational-models}

We used random disjoint and stratified splits of 70\% of the data for training and the remaining 30\% for testing. 
We used 10\% of the training set for validation.
According to \autoref{table:serp-layout-stats}, there are 2,443 samples containing DD ads, with 1,710 used for training and 733 for testing. Additionally, there are 2,647 samples with organic ads, comprising 1,853 for training and 794 for testing.
For both the SVM and $k$-NN models, we used the default parameters provided by the scikit-learn library in Python.
The RNN architecture had an input layer followed by a hidden GRU layer with hyperbolic tangent as activation function. 
The embedding size of the hidden layer is 150. 
The hidden layer is followed by a dropout operation ($q=0.25$),
then a fully connected layer followed by an output layer having a single neuron and sigmoid activation.
All models were trained using Adam optimizer with a learning rate $\eta=10^{-3}$ and default decay rates $\beta_1=0.9, \beta_2=0.999$.
Training was conducted in batches of 32 sequences each, for a maximum of 100 epochs.
Early stopping with a patience of 10 epochs was applied to prevent overfitting, 
using classification accuracy as the monitoring metric.

\subsection{Classification performance}
\label{sec:results}

\autoref{table:dd-ads-results-summary} and \autoref{table:organic-ads-results-summary} 
report the F$_1$ score (the harmonic average of Precision and Recall) and AUC score (area under the ROC curve), 
considering all possible combinations of our design parameters (model type, input representation, and trajectory duration). 
The best results were observed when using the first 5 seconds of the mouse trajectory as input 
with the GRU classifier: $F_1=93\%$ for organic ads and $F_1=73\%$ for DD ads.
By way of comparison, \citet{bruckner2021choice} also reported improved results when considering
a subset of the initial points in the mouse trajectories.

\begin{table}[!ht]
    \caption{
        Performance of baseline models,
        considering different input representations and mouse trajectory duration, 
        on SERPs containing any combination of DD ads.
    }
    \centering
    \small
    \begin{tabular}{lcc *5c}
        \toprule
        \textbf{Model type}  & \textbf{Duration (s)} & \textbf{F$_1$ score} & \textbf{AUC}\\
        \midrule
        SVM  & 5 & 0.67 & 0.71\\
        $k$-NN  & 5 & 0.57 & 0.57\\
        GRU  & 5 & 0.70 & 0.78\\
        \midrule
        SVM  & 10 & 0.63 & 0.67\\
        $k$-NN  & 10 & 0.54 & 0.55\\
        GRU  & 10 & 0.73 & 0.82\\
        \midrule
        SVM  & 15 & 0.59 & 0.59\\
        $k$-NN  & 15 & 0.54 & 0.54\\
        GRU  & 15 & 0.72 & 0.77\\
        \midrule
        SVM  & 20 & 0.60 & 0.60\\
        $k$-NN  & 20 & 0.55 & 0.55\\
        GRU  & 20 & 0.69 & 0.74\\
        \bottomrule
    \end{tabular}
    \label{table:dd-ads-results-summary}
\end{table}

\begin{table}[!ht]
    \caption{
        Performance of baseline models,
        considering different input representations and mouse trajectory duration, 
        on SERPs containing only organic ads.
    }
    \centering
    \small
    \begin{tabular}{lcc *5c}
        \toprule
        \textbf{Model type}  & \textbf{Duration (s)} & \textbf{F$_1$ score} & \textbf{AUC}\\
        \midrule
        SVM  & 5 & 0.89 & 0.94\\
        $k$-NN  & 5 & 0.86 & 0.84\\
        GRU  & 5 & 0.93 & 0.97\\
        \midrule
        SVM  & 10 & 0.92 & 0.95\\
        $k$-NN  & 10 & 0.85 & 0.82\\
        GRU  & 10 & 0.93 & 0.97\\
        \midrule
        SVM  & 15 & 0.88 & 0.90\\
        $k$-NN  & 15 & 0.77 & 0.76\\
        GRU  & 15 & 0.90 & 0.96\\
        \midrule
        SVM  & 20 & 0.86 & 0.91\\
        $k$-NN  & 20 & 0.71 & 0.71\\
        GRU  & 20 & 0.89 & 0.95\\
        \bottomrule
    \end{tabular}
    \label{table:organic-ads-results-summary}
\end{table}

\section{Discussion, limitations, and future work}
\label{sec:results}

\begin{figure}[!ht]
  \centering
  \includegraphics[width=\linewidth]{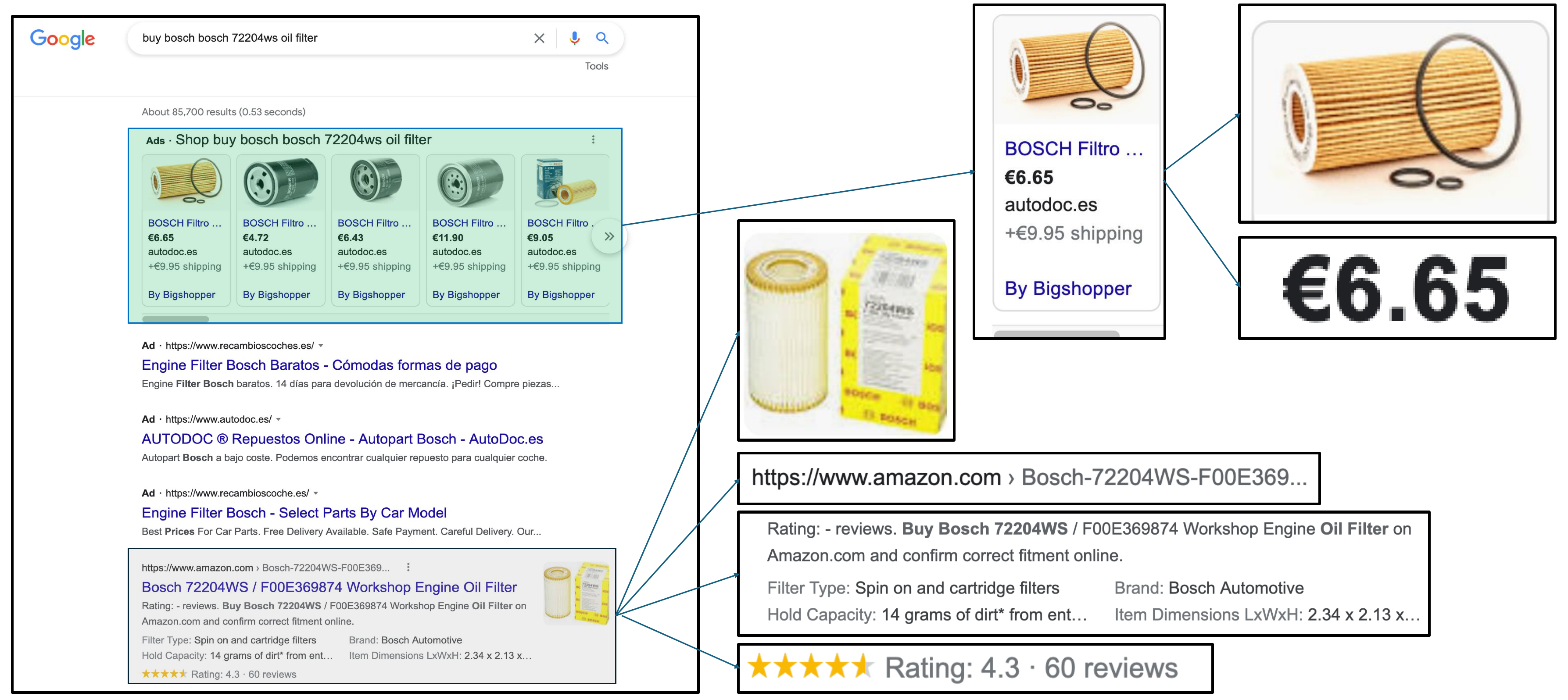}
  \caption{
    Examples of fine-grained AOIs with additional information. 
    For DD ads, images and price values can be extracted. 
    For organic items, the corresponding links, descriptions, ratings, and review information can also be extracted.
  }
  \label{fig:potential-aoi-example}
\end{figure}

Our dataset can be primarily used to assess user interaction with SERPs 
by predicting users' attention on ads or other page elements, 
either explicitly from eye-tracking data or implicitly from mouse cursor data. 
However, we anticipate many other interesting uses.
For example, the ads considered in our baseline experiments 
consist of various DOM elements, such as descriptions, prices, and images (see \autoref{fig:potential-aoi-example}). 
Predicting users' attention on these particular elements could be another interesting topic for future research. 
Extracting text from the ad descriptions can facilitate analyses of semantic relevance w.r.t the input query, 
as well as their correlation with visual attention. 
A similar approach can be applied to images in other SERP regions;
e.g., they can be further analyzed by using computer vision techniques.
Inspired by previous work~\cite{arapakis2020learning}, 
we have also provided additional materials (see \autoref{fig:representations}) 
that future studies can utilize to train Convolutional Neural Network (CNN) models, for example.

While typical computer mice are the most common and widely used devices for web browsing on desktop computers, 
one of our limitations is that we did not consider other types of input devices 
that can control the mouse cursor, such as trackpads on laptops.
We have not considered mobile devices either, 
since they do not provide fine-grained movement data.
Additionally, as shown in \autoref{fig:demographics-characteristics}, most participants preferred shopping on Amazon. Therefore, future work could replicate our study in e-commerce websites to further validate our findings. Another limitation of our work is that the search queries were given to the participants, following previous work~\cite{arapakis2014understanding, arapakis2016predicting, arapakis2020learning}.

In terms of model performance, in our baseline experiments 
we have observed $F_1$ scores of 93\% for organic ads and 73\% for DD ads,
which are higher than the results reported in the closest work to ours~\cite{arapakis2020learning}
that used ``the attentive cursor dataset''~\cite{leiva2020attentive}.
We trained the same GRU model with the same input representation 
(i.e., mouse position and bounding box of the related ad) 
on the attentive cursor dataset 
and observed $F_1$ scores of 56\% and 69\% for organic and DD ads, respectively.
This suggests that self-reported labels are noisier than fixation-based labels,
especially when predicting attention to organic ads.

Further, and unsurprisingly, our modeling results show that input representations are critical 
for ensuring a high classification performance.
Interestingly, we found that our GRU model (which did automatic feature extraction) 
led to better results than SVM and $k$-NN models, which used Mouse2Vec features. 
One reason might be that Mouse2Vec embeddings did not see many SERPs during model pre-training~\cite{zhang2024mouse2vec}.
Therefore, finetuning Mouse2Vec on our dataset may improve the results.

To ensure that sessions remained within a manageable timeframe, the duration of each trial was limited to 1 minute. 
Nevertheless, our results showed that a short duration of 5 or 10 seconds is sufficient to predict visual attention from mouse movements, 
in line with previous findings~\cite{bruckner2021choice}. 
Additionally, our data indicate that the more the time spent on a SERP, 
the lower the likelihood of revisiting previous elements, 
suggesting that the initial seconds of the trial are more critical for assessing visual attention. 
This aligns with previous findings by \citet{bruckner2021choice},
who noted that, for visual attention tasks, 
it was more convenient to consider the first few mouse movements 
instead of considering the full mouse trajectory.
For other tasks it may be interesting to consider other movement lengths
and even other input representations, such as the SERP screenshots
or the fine-grained segmentation of AOIs depicted in \autoref{fig:potential-aoi-example}.

\section{Conclusion}
\label{sec:conclusion}

We contribute a large-scale in-lab dataset to study coordinated mouse and eye movements on SERPs. 
The main differentiating factors of this dataset are: 
(1)~self-contained source HTML files;
(2)~fine-grained AOI identification;
(3)~SERP screenshots available in several configurations;
and (4)~eye-tracking data to construct objective ground-truth labels.
Additionally, we trained computational models to assess users' attention 
by using different representations of mouse movements and fixation-based labels.
We also have revisited previous research findings regarding how users interact with SERPs
We also have discussed the new insights derived from our dataset, 
which will be released in full upon the publication of this paper.

\begin{acks}
We thank Mateusz Dubiel for his feedback during the pilot experiments.
We also thank Hugo Barthelemy for helping us with the implementation of AOI extraction from HTML files.
This research is supported by the Horizon 2020 FET program of the European Union (grant CHIST-ERA-20-BCI-001)
and the European Innovation Council Pathfinder program (SYMBIOTIK project, grant 101071147).
\end{acks}




\end{document}